# Machine learning predicts early onset of fever from continuous physiological data of critically ill patients

Aditya Singh[1], Akram Mohammed[1] Lokesh Chinthala[1], Rishikesan Kamaleswaran[1*]

[1] University of Tennessee Health Science Center, Memphis, TN, USA

[*] Corresponding Author
Rishikesan Kamaleswaran, Ph.D.
50 N. Dunlap St, 4th Floor 491R,
Memphis, 38103, TN, USA
E: rkamales@uthsc.edu

## Abstract

**Background and Objective:** Fever can provide valuable information for diagnosis and prognosis of various diseases such as pneumonia, dengue, sepsis, etc., therefore, predicting fever early can help in the effectiveness of treatment options and expediting the treatment process. The aim of this study is to develop novel algorithms that can accurately predict fever onset in critically ill patients by applying machine learning technique on continuous physiological data.
**Methods:** We analyzed continuous physiological data collected every 5-minute from a cohort of over 200,000 critically ill patients admitted to an Intensive Care Unit (ICU) over a 2-year period. Each episode of fever (temperature≥38.0 Celsius) from the same patient were considered as an independent event, with separations of at least 24 hours. We extracted descriptive statistical features from six physiological data streams, including heart rate, respiration, systolic and diastolic blood pressure, mean arterial pressure, and oxygen saturation, and use these features to independently predict the onset of fever. Using a bootstrap aggregation method, we created a balanced dataset of 7,801 afebrile and febrile patients and analyzed features up to 4 hours before the fever onset.
**Results:** We applied XGBoost machine learning algorithm and achieved an average 5-fold cross-validated recall, precision, and F1-score of 95%, 97.5%, and 94%, respectively.
**Conclusions:** Supervised machine learning method can predict fever up to 4 hours before onset in critically ill patients with high recall, precision, and F1-score. This study demonstrates the viability of using machine learning to predict fever among hospitalized adults. The discovery of salient physiomarkers through machine learning and deep learning techniques has the potential to further accelerate the development and implementation of innovative care delivery protocols and strategies for medically vulnerable patients.

## Introduction

Up to half of the patients admitted in the adult intensive care units are observed with elevated core temperature (1–3). The magnitude of fever has been associated with increased mortality among adults, especially among post-operative patients (4, 5). Fever can be accompanied by a number of clinical and physiological observations, including elevated heart rate, tachypnea and low blood pressures (4). In critically ill patients, the presence of fever may prompt rapid work-ups, including for patients with infection or in traumatic brain injury (6). Especially among patients with central fever, rapid antipyretics are often administered to preserve brain function (7). Therefore, predicting the onset of fever can significantly enhance clinical decision making and prompt earlier therapy.

Patients admitted to the ICU are monitored by a host of medical devices. Data generated from those devices are largely discarded during the process of care. Recent work has suggested that features or 'physiomarkers' generated from continuous physiological data streams can predict the onset of physiological deterioration in children and adults (8–12). Increasing the frequency of those vitals, from ones captured hourly to every minute, has been shown to improve the predictive performance of such models (13). In this paper, we identify novel physiomarkers in the continuous data stream that predict the onset of fever, independent of the temperature recordings. We apply machine learning methods to generate time-frequency domain features that are clinically meaningful. Therefore, the model can assist in clinical decision support for selecting patients who are at risk for developing an imminent fever.

## Results

Of the 200,859 patient encounters from the eICU dataset (See methods for more details), we included 19,419 patients who had the temperature measurements at a 5-minute interval. Thirteen thousand and ninety-eight fever episodes resulted from 9,785 unique patient encounters and remaining patient encounters were used as the controls 4 hours before fever onset. A total of 7,801 afebrile patients were identified, however, to balance our dataset, we randomly selected an equal number (from a set of 13,098 cases) of cases and repeated this step 100 times and report the mean classification statistics.

### Feature Extraction

Figure 1 shows the frequency of all the 42 features generated from six physiologic signals that are used to build the machine learning models. The X-axis represents the F-score value of each of the feature extracted and Y-axis represents the names of the features from the physiological signal. Feature selection was performed to reduce the number of features, and the reduced feature set was fed into each of the classifiers. We extracted feature importance from feature set using XGBoost classifier and identified the maximum oxygen saturation of time-series physiologic variable as the most important feature, followed by standard deviation and mean feature generated from oxygen saturation of the time-series data.

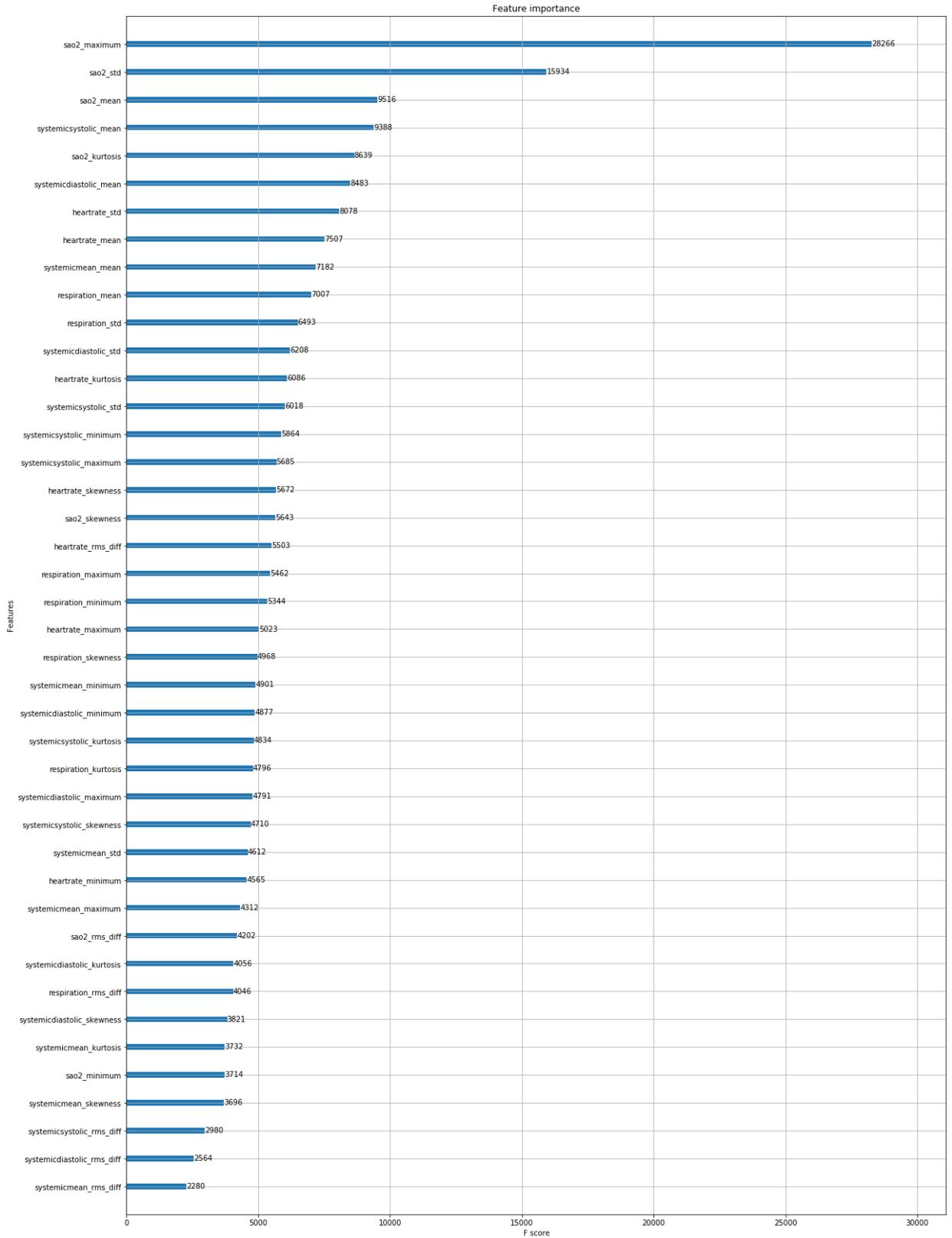

Figure 1: Features derived from physiologic signals up to two hours prior to fever episode. sao2=oxygen saturation; rms_diff=root mean square difference

Figure 2 shows the difference between febrile and afebrile class for each of the discriminatory features. Minimum mean blood pressure, kurtosis respiratory rate, kurtosis heart rate, and minimum respiratory rate

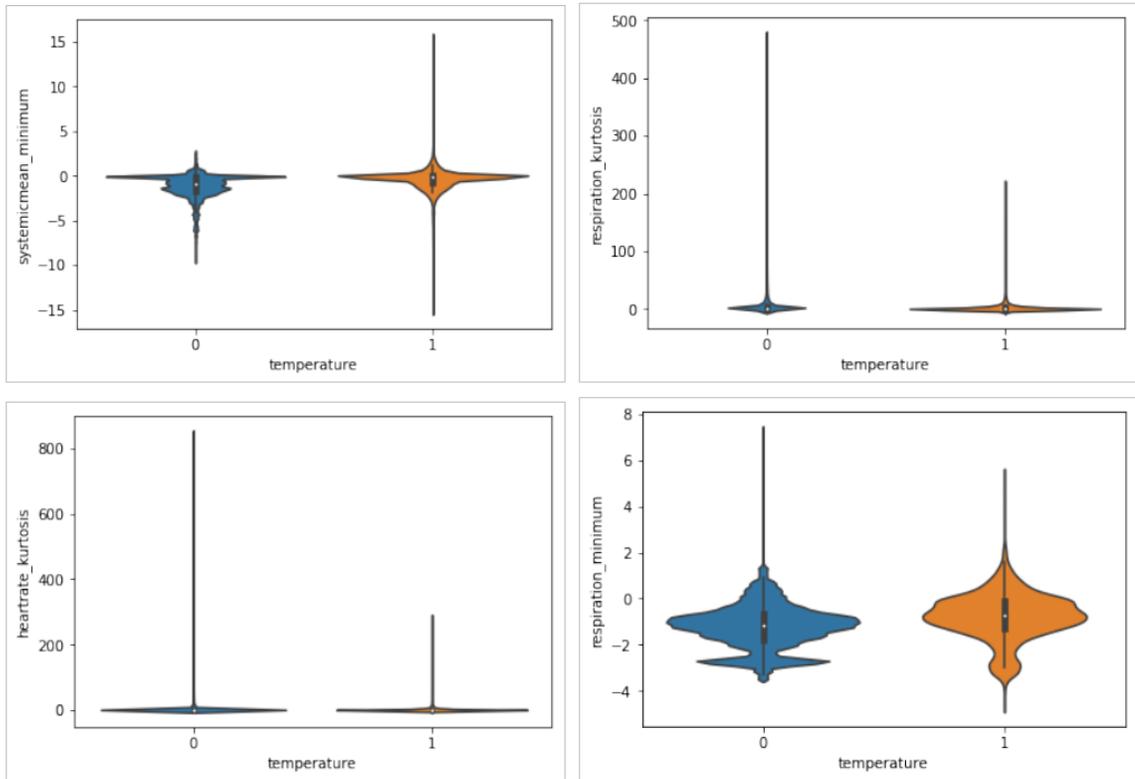

Figure 2: Exploring data with violin plots: The most discriminatory extracted features from the 42-feature set

**Predictive Power of the Models for fever onset**
Five-fold cross-validation model was developed using XGBClassifier. The average cross-validated recall, precision, and F1-score from on training data are 95%, 97.5%, and 94%, respectively, 4 hours before fever onset. The average recall on the 20% test set are 82.4%. The analysis was conducted by excluding predictors from 4 hrs. window before fever onset.

## Methods
**Dataset**
We used continuous physiological data collected every 5-minute from 200,859 patient encounters from 139,365 critically ill patients admitted to an Intensive Care Unit (ICU) over a 2-year period in 208 hospitals from the eICU Collaborative Research Database (14). Of the 200,859 patient encounters, we excluded 181,440 patient encounters who did not have the temperature measurements. The missing data for all the variables other than the temperature were filled with the median value. We collected the following physiological signals: heart rate, respiration, central venous pressure, systolic and diastolic blood pressure, mean arterial pressure, and oxygen saturation and set a data availability threshold of at least 50% and therefore, removed central venous pressure. The percentage of data availability is given in Table 1.

Table 1: Percentage data availability for physiological variables

| Physiological signal | % Data availability |
|---|---|
| Temperature | 100.00 |
| Heart rate | 99.99 |
| Oxygen saturation | 99.03 |
| Respiratory rate | 94.77 |
| Mean arterial pressure | 56.15 |
| Systolic blood pressure | 55.96 |
| Diastolic blood pressure | 55.95 |
| Central venous pressure | 45.32 |

**Feature extraction**
We extracted seven descriptive features, including standard deviation, mean, minimum, maximum, kurtosis, root mean square difference, skewness from six physiological data streams, including heart rate, respiration, systolic and diastolic blood pressure, mean arterial pressure, and oxygen saturation, a total of 42 features. Each episode of fever (temperature $>>=38.0$ Celsius) from the same patient were considered as an independent event, with separations of at least 24 hours.

**Machine learning**
We used XGBoost machine learning method, which is an efficient, portable, optimized distributed gradient boosting library (15). XGBoost provides a parallel tree boosting that solve many data science problems in a fast and accurate way. The distributed environment allows the efficient running of code to solve problems due to data size. Using a bootstrap aggregation method, we created a balanced dataset of 7,801 afebrile patients and febrile (from 13,098) and analyzed features up to 4 hours before the fever onset. The data is first partitioned into the train (80%) and test (20%) datasets. The train set is used for cross-building validated models, while the test set was used to perform the model validation.

**Hyper parameterization**
We used the grid search for finding the best hyperparameter and trained the models iteratively with different sets of controls and the same set of cases (Cluster-Based Over Sampling) and partially fit the model (XG Boost). We have used the following hyperparameters for XGBoost: subsample of 0.25 to make it more conservative. For Booster, we explored gbtree and dart and eventually gbtree was selected. The learning rate of 0.3 was used which shrinks the feature weights to make the boosting process more conservative. The following maximum depths were checked: 12, 10, 8, 6, 4, 3 and 2. The objective of 'reg: linear' was selected, whereas, the number of rounds for boosting of 100 was used.

**Cross-Validation**
Models were developed from distinct time intervals. For model selection and accuracy estimation, we used 5-fold cross-validation (16). This technique divides data (stratified partitioning) into five equal and discrete folds and uses four folds for the model generation, while predictions are generated and evaluated using the remaining one-fold. This step is subsequently repeated five times, so each fold is tested against the other four folds. We further ran each of these 5-fold cross-validation models 100 times by shuffling the data in each iteration and averaged the performance metrics from all iterations to reduce bias.

**Statistical Analysis and machine learning framework**
Python scikit-learn machine learning library (17) was used for calculating descriptive statistical measures, feature selection and building machine learning classifiers.

# Discussion and Conclusions
One of the important limitations of machine learning use in clinical decision making is that machine learning algorithms are less intuitive compared to the statistical method. Supervised machine learning methods can predict fever up to 4 hours before onset in critically ill patients with high recall, precision, and F1-score. This study demonstrates the viability of using machine learning to predict fever among hospitalized adults. The discovery of salient physiomarkers through machine learning technique has the potential to further accelerate the development and implementation of innovative care delivery protocols and strategies for medically vulnerable patients.

Though fever is itself is not a major challenge, but it can lead to various major complications. Therefore, predicting early fever onset is critical. The ability of predictive algorithms to identify patients at high risk for fever by using routinely collected physiological data can provide important early warnings of impending physiological deterioration. Such information can help in clinical decision making and may even eventually be useful in guiding early goal-directed therapy. In this study, we demonstrated that the machine learning-based prediction models could accurately distinguish fever patients up to 4 hours prior to the onset. The classifiers and the selected physiologic features may facilitate accurate, unbiased fever diagnosis and prediction and effective treatment, ultimately improving prognosis.

The selection of relevant features involved in fever and control remains a challenge. Therefore, it is important to find a subset of physiologic features that are sufficiently informative to distinguish between febrile and afebrile patients. To extract useful information from these patients' continuous physiologic data and to reduce dimensionality, feature-selection algorithms were systematically investigated in this study. As we have demonstrated in the results, selecting subsets of physiological variables based on the availability of the data allowed for the high performance of our classification models. Salient physiomarkers (such as descriptive statistical features) derived from the physiological signals such as oxygen saturation, blood pressure, heart rate, and respiratory rate, may precede various complications in febrile patients as suggested by the results in this study. Though these machine learning physiomarkers are neither observable by physicians nor are they readily interpretable; but the advantage could be primarily towards the early prediction of health deterioration and alerting health care providers of that fact. Further research is needed to understand how to use these alerts to guide personalized care of patients with fever. The results indicate that among all classifiers, XGBoost classifier provides a good balance between the precision and recall even at 4 hours prediction window before the fever onset.

In conclusion, we showed, as a proof of principle, that machine learning can accurately predict the fever in ICU patients with up to 4 hours before onset. This finding is significant because it may optimize the early recognition of serious disease complications and allow for the implementation of early interventions.

## Acknowledgment
We would like to acknowledge the physionet.org and Google summer of code program.

## Author contributions
RK conceptualized and designed the study, participated in the editing of the manuscript and critically reviewed the manuscript for important intellectual content. AS carried out the data analysis and participated in the editing of the manuscript. AM carried out the initial data preprocessing, data analyses and participated in the drafting of the manuscript. LC critically reviewed the manuscript for important intellectual content.

## Conflict of interest
The authors declare no competing interest.